\documentclass[useAMS,usenatbib]{mn2e}
\usepackage{tabularx}
\usepackage{xspace}
\usepackage{graphicx}
\newcommand{\ignore}[1]{}
\providecommand{\ao}{}

\renewcommand{\ao}{adaptive optics (AO)\renewcommand{\ao}{AO\xspace}\renewcommand{\Ao}{AO\xspace}\xspace}
\newcommand{\Ao}{Adaptive optics (AO)\renewcommand{\ao}{AO\xspace}\renewcommand{\Ao}{AO\xspace}\xspace}
\newcommand{\wfs}{wavefront sensor (WFS)\renewcommand{\wfs}{WFS\xspace}\renewcommand{\wfss}{WFSs\xspace}\xspace}
\newcommand{\wfss}{wavefront sensors (WFSs)\renewcommand{\wfs}{WFS\xspace}\renewcommand{\wfss}{WFSs\xspace}\xspace}
\newcommand{\shwfs}{Shack-Hartmann \wfs (SHWFS)\renewcommand{\shwfs}{SHWFS\xspace}\xspace}
\newcommand{\dm}{deformable mirror (DM)\renewcommand{\dm}{DM\xspace}\renewcommand{\dms}{DMs\xspace}\renewcommand{\Dms}{DMs\xspace}\renewcommand{\Dm}{DM\xspace}\xspace}
\newcommand{\dms}{deformable mirrors (DMs)\renewcommand{\dm}{DM\xspace}\renewcommand{\dms}{DMs\xspace}\renewcommand{\Dms}{DMs\xspace}\renewcommand{\Dm}{DM\xspace}\xspace}
\newcommand{\Dms}{Deformable mirrors (DMs)\renewcommand{\dm}{DM\xspace}\renewcommand{\dms}{DMs\xspace}\renewcommand{\Dms}{DMs\xspace}\renewcommand{\Dm}{DM\xspace}\xspace}
\newcommand{\Dm}{Deformable mirror (DM)\renewcommand{\dm}{DM\xspace}\renewcommand{\dms}{DMs\xspace}\renewcommand{\Dms}{DMs\xspace}\renewcommand{\Dm}{DM\xspace}\xspace}

\newcommand{\lqg}{linear-quadratic-gaussian (LQG)\renewcommand{\lqg}{LQG\xspace}\xspace}
\newcommand{\shs}{Shack-Hartmann sensor (SHS)\renewcommand{\shs}{SHS\xspace}\renewcommand{\shss}{SHSs\xspace}\xspace}
\newcommand{\shss}{Shack-Hartmann sensors (SHSs)\renewcommand{\shs}{SHS\xspace}\renewcommand{\shss}{SHSs\xspace}\xspace}
\newcommand{\lgs}{laser guide star (LGS)\renewcommand{\lgs}{LGS\xspace}\renewcommand{\lgs}{LGS\xspace}\renewcommand{\lgss}{LGSs\xspace}\xspace}
\newcommand{\lgss}{laser guide stars (LGSs)\renewcommand{\lgs}{LGS\xspace}\renewcommand{\lgs}{LGS\xspace}\renewcommand{\lgss}{LGSs\xspace}\xspace}
\newcommand{\Lgs}{Laser guide star (LGS)\renewcommand{\lgs}{LGS\xspace}\renewcommand{\Lgs}{LGS\xspace}\renewcommand{\lgss}{LGSs\xspace}\xspace}

\newcommand{\Ngs}{Natural guide star (NGS)\renewcommand{\ngs}{NGS\xspace}\renewcommand{\Ngs}{NGS\xspace}\renewcommand{\ngss}{NGSs\xspace}\xspace}
\newcommand{\ngs}{natural guide star (NGS)\renewcommand{\ngs}{NGS\xspace}\renewcommand{\Ngs}{NGS\xspace}\renewcommand{\ngss}{NGSs\xspace}\xspace}
\newcommand{\ngss}{natural guide stars (NGSs)\renewcommand{\ngs}{NGS\xspace}\renewcommand{\Ngs}{NGS\xspace}\renewcommand{\ngss}{NGSs\xspace}\xspace}
\newcommand{\mems}{Micro-Electro-Mechanical Systems (MEMS)\renewcommand{\mems}{MEMS\xspace}\xspace}
\newcommand{\snr}{signal to noise ratio (SNR)\renewcommand{\snr}{SNR\xspace}\xspace}
\newcommand{\Moao}{Multi-object \ao (MOAO)\renewcommand{\moao}{MOAO\xspace}\renewcommand{\Moao}{MOAO\xspace}\xspace}
\newcommand{\moao}{multi-object \ao (MOAO)\renewcommand{\moao}{MOAO\xspace}\renewcommand{\Moao}{MOAO\xspace}\xspace}
\newcommand{\mcao}{multi-conjugate adaptive optics (MCAO)\renewcommand{\mcao}{MCAO\xspace}\xspace}
\newcommand{\ltao}{laser tomographic \ao (LTAO)\renewcommand{\ltao}{LTAO\xspace}\xspace}
\newcommand{\cpu}{central processing unit (CPU)\renewcommand{\cpu}{CPU\xspace}\renewcommand{\cpus}{CPUs\xspace}\xspace}
\newcommand{\cpus}{central processing units (CPUs)\renewcommand{\cpu}{CPU\xspace}\renewcommand{\cpus}{CPUs\xspace}\xspace}
\newcommand{\psf}{point spread function (PSF)\renewcommand{\psf}{PSF\xspace}\renewcommand{\psfs}{PSFs\xspace}\renewcommand{\Psf}{PSF\xspace}\xspace}
\newcommand{\psfs}{point spread functions (PSFs)\renewcommand{\psf}{PSF\xspace}\renewcommand{\psfs}{PSFs\xspace}\renewcommand{\Psf}{PSF\xspace}\xspace}
\newcommand{\Psf}{Point spread function (PSF)\renewcommand{\psf}{PSF\xspace}\renewcommand{\psfs}{PSFs\xspace}\renewcommand{\Psf}{PSF\xspace}\xspace}
\newcommand{\fpga}{field programmable gate array (FPGA)\renewcommand{\fpga}{FPGA\xspace}\renewcommand{\fpgas}{FPGAs\xspace}\xspace}
\newcommand{\fpgas}{field programmable gate arrays (FPGAs)\renewcommand{\fpga}{FPGA\xspace}\renewcommand{\fpgas}{FPGAs\xspace}\xspace}
\newcommand{\sor}{successive over-relaxation (SOR)\renewcommand{\sor}{SOR\xspace}\xspace}
\newcommand{\fdpcg}{Fourier domain pre-conditioned gradient (FDPCG)\renewcommand{\fdpcg}{FDPCG\xspace}\xspace}
\newcommand{\map}{maximum a-posteriori (MAP)\renewcommand{\map}{MAP\xspace}\xspace}

\newcommand{\elt}{Extremely Large Telescope (ELT)\renewcommand{\elt}{ELT\xspace}\renewcommand{\elts}{ELTs\xspace}\renewcommand{\eelt}{European ELT (E-ELT)\renewcommand{\eelt}{E-ELT\xspace}\xspace}\xspace}

\newcommand{\elts}{Extremely Large Telescopes (ELTs)\renewcommand{\elt}{ELT\xspace}\renewcommand{\elts}{ELTs\xspace}\renewcommand{\eelt}{European ELT (E-ELT)\renewcommand{\eelt}{E-ELT\xspace}\xspace}\xspace}

\newcommand{\eelt}{European Extremely Large Telescope (E-ELT)\renewcommand{\eelt}{E-ELT\xspace}\renewcommand{\elt}{ELT\xspace}\renewcommand{\elts}{ELTs\xspace}\xspace}

\newcommand{\dugall}{Durham University generalised adaptive optics laser laboratory (DUGALL)\renewcommand{\dugall}{DUGALL\xspace}\xspace}
\newcommand{\fwhm}{full-width at half-maximum (FWHM)\renewcommand{\fwhm}{FWHM\xspace}\xspace}
\newcommand{\wht}{William Herschel Telescope (WHT)\renewcommand{\wht}{WHT\xspace}\xspace}
\newcommand{\emccd}{electron multiplying CCD (EMCCD)\renewcommand{\emccd}{EMCCD\xspace}\renewcommand{\emccds}{EMCCDs\xspace}\xspace}
\newcommand{\emccds}{electron multiplying CCDs (EMCCDs)\renewcommand{\emccd}{EMCCD\xspace}\renewcommand{\emccds}{EMCCDs\xspace}\xspace}
\newcommand{\dasp}{Durham \ao simulation platform (DASP)\renewcommand{\dasp}{DASP\xspace}\renewcommand{\thedasp}{DASP\xspace}\renewcommand{\Thedasp}{DASP\xspace}\xspace}
\newcommand{\thedasp}{the Durham \ao simulation platform (DASP)\renewcommand{\dasp}{DASP\xspace}\renewcommand{\thedasp}{DASP\xspace}\renewcommand{\Thedasp}{DASP\xspace}\xspace}
\newcommand{\Thedasp}{The Durham \ao simulation platform (DASP)\renewcommand{\dasp}{DASP\xspace}\renewcommand{\thedasp}{DASP\xspace}\renewcommand{\Thedasp}{DASP\xspace}\xspace}
\newcommand{\mpi}{Message Passing Interface (MPI)\renewcommand{\mpi}{MPI\xspace}\xspace}
\newcommand{\smp}{symmetric multi-processing (SMP)\renewcommand{\smp}{SMP\xspace}\xspace}
\newcommand{\svd}{singular value decomposition (SVD)\renewcommand{\svd}{SVD\xspace}\xspace}
\newcommand{\gpu}{graphics processing unit (GPU)\renewcommand{\gpu}{GPU\xspace}\renewcommand{\gpus}{GPUs\xspace}\xspace}
\newcommand{\gpus}{graphics processing units (GPUs)\renewcommand{\gpu}{GPU\xspace}\renewcommand{\gpus}{GPUs\xspace}\xspace}
\newcommand{\fft}{fast Fourier transform (FFT)\renewcommand{\fft}{FFT\xspace}\xspace}
\newcommand{\ifu}{integral field unit (IFU)\renewcommand{\ifu}{IFU\xspace}\xspace}
\newcommand{\darc}{the Durham \ao real-time controller (DARC)\renewcommand{\darc}{DARC\xspace}\renewcommand{\Darc}{DARC\xspace}\xspace}
\newcommand{\Darc}{The Durham \ao real-time controller (DARC)\renewcommand{\darc}{DARC\xspace}\renewcommand{\Darc}{DARC\xspace}\xspace}
\newcommand{\cots}{commercial off-the-shelf (COTS)\renewcommand{\cots}{COTS\xspace}\xspace}
\newcommand{\rtcp}{real-time control pipeline (RTCP)\renewcommand{\rtcp}{RTCP\xspace}\xspace}
\newcommand{\rms}{root-mean-square (RMS)\renewcommand{\rms}{RMS\xspace}\xspace}
\newcommand{\sFPDP}{serial Front Panel Data Port (sFPDP)\renewcommand{\sFPDP}{sFPDP\xspace}\xspace}
\newcommand{\wpu}{wavefront processing unit (WPU)\renewcommand{\wpu}{WPU\xspace}\xspace}

\newcommand{\rtcs}{real-time control system (RTCS)\renewcommand{\rtcs}{RTCS\xspace}\renewcommand{\rtcss}{RTCSs\xspace}\xspace}
\newcommand{\rtcss}{real-time control systems (RTCSs)\renewcommand{\rtcs}{RTCS\xspace}\renewcommand{\rtcss}{RTCSs\xspace}\xspace}
\newcommand{\eso}{European Southern Observatory (ESO)\renewcommand{\eso}{ESO\xspace}\renewcommand{\theeso}{ESO\xspace}\xspace}
\newcommand{\theeso}{\renewcommand{\theeso}{ESO\xspace}the \eso}
\newcommand{\scao}{single conjugate \ao (SCAO)\renewcommand{\scao}{SCAO\xspace}\renewcommand{\Scao}{SCAO\xspace}\xspace}
\newcommand{\Scao}{Single conjugate \ao (SCAO)\renewcommand{\scao}{SCAO\xspace}\renewcommand{\Scao}{SCAO\xspace}\xspace}
\newcommand{\glao}{ground layer \ao (GLAO)\renewcommand{\glao}{GLAO\xspace}\xspace}
\newcommand{\eagle}{ELT Adaptive optics for GaLaxy Evolution (EAGLE)\renewcommand{\eagle}{EAGLE\xspace}\xspace}
\newcommand{\maory}{multi-conjugate \ao relay for the \eelt (MAORY)\renewcommand{\maory}{MAORY\xspace}\xspace}
\newcommand{\muse}{Multi Unit Spectroscopic Explorer (MUSE)\renewcommand{\muse}{MUSE\xspace}\xspace}
\newcommand{\vlt}{Very Large Telescope (VLT)\renewcommand{\vlt}{VLT\xspace}\xspace}

\newcommand{\tmt}{Thirty Metre Telescope (TMT)\renewcommand{\tmt}{TMT\xspace}\xspace}
\newcommand{\xao}{eXtreme \ao (XAO)\renewcommand{\xao}{XAO\xspace}\xspace}

\newcommand{\vla}{Very Large Array (VLA)\renewcommand{\vla}{VLA\xspace}\xspace}
\newcommand{\jwst}{{\em James Webb Space Telescope} \citep[JWST,][]{jwst}\renewcommand{\jwst}{{\em JWST}\xspace}\xspace}
\newcommand{\hst}{{\em Hubble Space Telescope (HST)}\renewcommand{\hst}{{\em HST}\xspace}\xspace}
\newcommand{\ifss}{integral-field spectrographs (IFSs)\renewcommand{\ifss}{IFSs\xspace}\renewcommand{\ifs}{IFS\xspace}\xspace}
\newcommand{\ifs}{integral-field spectrograph (IFS)\renewcommand{\ifss}{IFSs\xspace}\renewcommand{\ifs}{IFS\xspace}\xspace}
\newcommand{\ifus}{integral field units (IFUs)\renewcommand{\ifus}{IFUs\xspace}\xspace}
\newcommand{\mos}{multi-object spectrograph (MOS)\renewcommand{\mos}{MOS\xspace}\xspace}
\newcommand{\goodss}{Great Observatories Origins Deep Survey (GOODS)-S\renewcommand{\goodss}{GOODS-S\xspace}\xspace}
\newcommand{\goods}{Great Observatories Origins Deep Survey (GOODS)\renewcommand{\goods}{GOODS\xspace}\xspace}
\newcommand{\scmos}{scientific CMOS (sCMOS)\renewcommand{\scmos}{sCMOS\xspace}\xspace}
\newcommand{\aof}{Adaptive Optics Facility (AOF)\renewcommand{\aof}{AOF\xspace}\xspace}
\newcommand{\dsp}{digital signal processor (DSP)\renewcommand{\dsp}{DSP\xspace}\renewcommand{\dsps}{DSPs\xspace}\xspace}
\newcommand{\dsps}{digital signal processors (DSPs)\renewcommand{\dsp}{DSP\xspace}\renewcommand{\dsps}{DSPs\xspace}\xspace}
\newcommand{\capi}{Coherent Accelerator Processor Interface (CAPI)\renewcommand{\capi}{CAPI\xspace}\xspace}
\newcommand{\qe}{quantum efficiency (QE)\renewcommand{\qe}{QE\xspace}\xspace}
\newcommand{\numa}{non-uniform memory access (NUMA)\renewcommand{\numa}{NUMA\xspace}\xspace}
\newcommand{\uav}{unmanned aerial vehicle (UAV)\renewcommand{\uav}{UAV\xspace}\renewcommand{\uavs}{UAVs\xspace}\xspace}
\newcommand{\uavs}{unmanned aerial vehicles (UAVs)\renewcommand{\uav}{UAV\xspace}\renewcommand{\uavs}{UAVs\xspace}\xspace}

\usepackage{amssymb}

\title[E-ELT MCAO performance modelling]{Monte-Carlo
  modelling of multi-conjugate adaptive optics performance on the
  European Extremely Large Telescope}

\author[A.\ G.\ Basden et al.]{A.\ G.\ Basden$^{1}$\thanks{E-mail:
    a.g.basden@durham.ac.uk (AGB)}\\
$^{1}$Department of Physics, South Road, Durham, DH1 3LE, UK}

\begin{document}
\maketitle

\begin{abstract}
The performance of a wide-field adaptive optics system depends on
input design parameters.  Here we investigate the performance of a
multi-conjugate adaptive optics system design for the European
Extremely Large Telescope, using an end-to-end Monte-Carlo adaptive
optics simulation tool, DASP.  We consider parameters such as the
number of laser guide stars, sodium layer depth, wavefront sensor
pixel scale, number of deformable mirrors, mirror conjugation and
actuator pitch.  We provide potential areas where costs savings can be
made, and investigate trade-offs between performance and cost.  We
conclude that a 6 laser guide star system using 3 DMs seems to be a
sweet spot for performance and cost compromise.
\end{abstract}


\section{Introduction}
\label{sect:intro}
The forthcoming \elts \citep{eelt,tmt,gmt} all rely on \ao systems
\citep{adaptiveoptics} to provide atmospheric turbulence compensation
allowing scientific goals requiring high resolution imaging and
spectroscopy to be met.  The design of these \ao systems requires
extensive modelling and simulation to enable performance estimates to
be made, and to explore relevant parameter spaces.  Current modelling
tools fall into two broad categories: analytical models, and
Monte-Carlo simulations.  Monte-Carlo simulations, while generally
computationally expensive for \elt-scale models, have the ability to
deliver high fidelity performance estimates, and include non-linear
effects, and as much system detail as is necessary (at the expense of
computational requirements).

Here, we use \thedasp \citep{basden5,basden11} to model expected \ao
performance for a \mcao instrument on the 39~m \eelt.  \dasp is a
Monte-Carlo end-to-end simulation tool, that includes models of the
atmosphere, telescope, wavefront sensors, deformable mirrors, \ao
real-time control system, and performance characterisation via
generation of science \psf images.  We investigate \ao performance as
a function of the number of \lgss \citep{laserguidestar}, the number
of \dms, \dm actuator pitch and conjugate height, and explore
performance across the telescope field of view.  We also investigate
the degree of elongation of \lgss, which is determined by sodium layer
depth in the mesosphere, and the impact of \wfs pixel scale on \ao
performance under different signal-to-noise regimes.  Our findings can
be used to aid instrument design decisions and to estimate expected
\ao performance for future instruments, and are complementary to
results from other modelling tools for \elt \mcao instrumentation, for
example \citet{2014SPIE.9148E..6FA,miskaltao}.

In \S2 we describe our input models and the explored
parameter space and in \S3 we present the performance estimates
obtained.  We conclude in \S4.

\section{Modelling of an E-ELT MCAO system}
We use \dasp to investigate different configurations for a \mcao
system on the \eelt.  In the study presented here, we usually consider only
the use of \lgss, in order to simplify our results.  We assume that
the tip-tilt signal from the \lgss is valid, so that \ngs measurements
are not necessary.  However, we also present results when using low
order \ngss for tip-tilt correction (and ignore tip-tilt signal from
the \lgs when doing so).  A previous study has investigated many different
\ngs asterisms for a \moao instrument on the \eelt \citep{basden17},
and so we do not seek to perform such a study here.  We note that the
assumption of a valid \lgs tip-tilt signal can be both pessimistic and
optimistic, depending on \ngs asterism, and \lgs asterism diameter,
since the \lgs locations are typically close to the edge of the field
of view, while the \ngs locations can be spread over the field.

Our key \ao performance metric is H-band Strehl ratio (1650~nm), as
this is an easily understandable measurement and of particular
relevance for imaging cameras, which are typically used behind \mcao
systems.

The \eelt design includes a \dm that is part of the telescope
structure (the fourth mirror in the optical train).  We therefore use
this as the ground layer conjugated \dm in our modelling, though it is
likely that this \dm will actually be conjugated a few hundred meters
away from ground level, an effect that we investigate.  

\subsection{Simulation model details}
The \eelt design has four sodium laser launch locations, equally
spaced around the telescope just beyond the edge of the telescope
aperture, i.e.\ the lasers are side-launched, rather than centre
launched, so that fratricide effects are irrelevant
\citep{2013aoel.confE..58O}.  At each launch location, up to two
lasers can be launched, with independent pointing possible.  In the
model that we use here, the launch locations are placed 22~m from the
centre of the telescope aperture.

We model the atmosphere using a standard \eso 35 layer turbulence
profile \citep{35layer} with an outer scale of 25~m and a Fried's
parameter of 13.5~cm at zenith.  Investigations into \ao performance with
variations of this atmospheric profile have been studied previously
\citep{basden17}, so we do not consider this further here: it is
important to realise that the performances derived here are relevant
for one atmospheric model only, and that the \ao performance will
differ under other atmospheric conditions.  Our simulations are
performed at 30$^\circ$ from zenith.

\begin{figure}
\includegraphics[width=\linewidth]{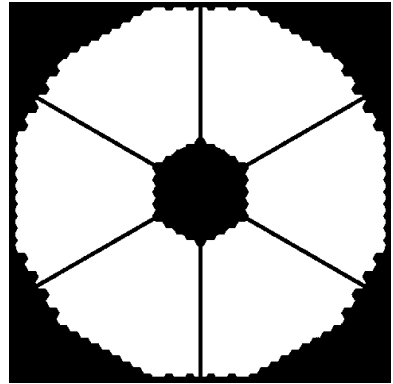}
\caption{A representitive E-ELT pupil function, for a line of sight
  1.5~arcminutes off-axis.  The hexagonal pattern of the segmented
  primary mirror is evident, and a slight vignetting by M4 is seen
  around the central obscuration (with the effect becoming more
  pronounced further off-axis).  A slight defocusing can also be seen,
due to the different conjugate heights of different mirrors in the
optical train.}
\label{fig:pupil}
\end{figure}

We assume $74\times74$ sub-apertures for each \lgs, and a telescope
diameter of 38.55~m, with a central obscuration diameter of 11~m.  We
model a direction dependant telescope pupil function, since for the
\eelt, the observed central obscuration changes across the field of
view.  Our model includes the hexagonal pattern of the segmented
mirrors, and telescope support structures (spiders), with a
representative pupil function shown in Fig.~\ref{fig:pupil}.  

Our \lgs asterism is arranged regularly on a circle, the diameter of
which we investigate.  The \lgs spots are elongated to model a sodium
layer at 90~km with a \fwhm between 5--20~km, and we include the cone
effect (or focal anisoplanatism, due to the finite distance) in our
simulations.  The \lgs \psfs are atmospherically broadened to
1~arcsecond, such that the minimum spot size has a 1~arcsecond \fwhm
along the \lgs axis.  The \lgs photon return is generally assumed to
be in the high-light regime (we use $10^6$ detected photons per
sub-aperture per frame) unless otherwise stated.  However, we also
investigate more realistic photon flux: the flux from the \eso
Wendelstein \lgs unit returns between 5--21 million photons per second
per m$^2$ \citep{caliaPrivate}, depending on location on the sky.
With a 500~Hz frame rate and 0.5~m sub-apertures, we can therefore
expect between 2500--10000 photons per sub-aperture per frame, with
additional reductions due to throughput losses and detector quantum
efficiency.  We include photon shot noise in our simulations.

We model detector readout noise at 0, 0.1 and 1 electrons rms per
pixel, corresponding to typical levels from a noiseless detector (the
default case), an \emccd, and a \scmos detector respectively.  For
simplicity, we assume that all pixels have the same rms readout noise.
This is not the case for \scmos technology meaning that our results
will be slightly optimistic.  However, this assumption has been
explored elsewhere \citep{basden19}.

We include a wavefront slope linearisation algorithm using a look-up
table to reduce the effect of non-linearity in the \wfs measurements.
The \lgs wavelength is 589~nm.  We measure science \psf performance at
1.65~$\mu$m.

We perform tomographic wavefront reconstruction at the conjugate
heights of the \mcao \dms using a minimum mean square error (MMSE)
algorithm \citep{map} with a Laplacian regularisation to approximate
wavefront phase covariance.  We assume a \wfs frame rate of 500~Hz,
and ensure that the science \psfs are well averaged.

The \dms are modelled using a cubic spline interpolation function,
which uses given actuator heights and positions to compute a surface
map of the \dm.  The ground layer conjugate \dm (M4) has $75\times75$
actuators, while the pitch of the higher layer conjugate \dms is
explored (with the number of required actuators depending on conjugate
height, pitch and field of view).  We do not consider \dm
imperfections, as this has been studied previously \citep{basden15}.

Unless otherwise stated, we use the following default parameters in
the results presented here: 6 \lgss evenly spaced around a 2~arcminute
diameter circle, at 90~km with a sodium layer \fwhm of 5~km, 3 \dms
conjugated to 0~km, 4~km and 12.7~km, with a 1~m actuator pitch (when
propagated to the telescope pupil) for the higher \dms, and a 0.52~m
actuator pitch for the ground layer \dm (equal to the sub-aperture
pitch).  We note that the assumption of a 5~km \fwhm sodium layer is
optimistic, however this was chosen to alleviate spot truncation
\citep[which is an issue studied elsewhere, e.g.\ ][]{2011aoel.confE..67V} when using
our default \lgs pixel scale of 0.23~arcsec/pixel (chosen to reduce
the computational complexity of our simulations).  A 10~km width is
more typical, while a 20~km width is considered pessimistic.  We also
note that the \dm conjugate heights are chosen to match tentative
designs for the first \eelt \mcao system, and also note that they are
similar to the GeMS system on the Gemini South telescope
\citep{2012SPIE.8447E..0IRshort}.  We use $16\times16$ pixels per
sub-aperture.  Results are presented on-axis, except where stated
otherwise.

Typically, we run our simulations for 5000 Monte-Carlo iterations,
representing 10~s of telescope time, which is sufficient to obtain a
well averaged \psf.  The uncertainties in our results due to
Monte-Carlo randomness are at the 1\% level, which we have verified
using a suite of separate Monte-Carlo runs.

\section{Predicted performance and uncertainties for E-ELT MCAO}
\begin{figure}
\includegraphics[width=\linewidth]{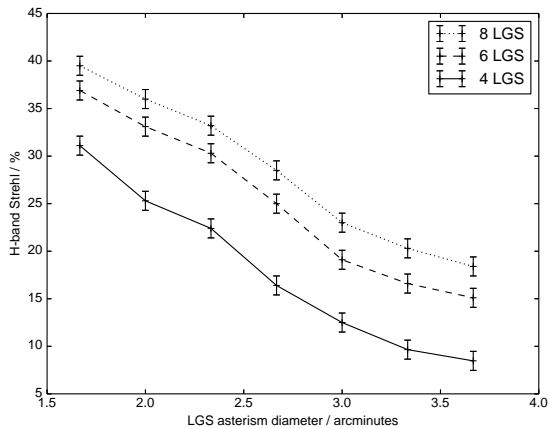}
\caption{A figure showing on-axis MCAO performance as a function of
  LGS asterism diameter.  The different curves are for different
  numbers of LGS, as given in the legend.}
\label{fig:nlgs}
\end{figure}

A key cost driver for \ao instruments on the \eelt is the number of
\lgss required.  Fig.~\ref{fig:nlgs} shows predicted \ao performance
at the centre of the field of view as a function of \lgs asterism
diameter, using different numbers of \lgs.  It can be seen that, as
expected, performance increases with the number of guide stars.
However, the gain in performance moving from 6 to 8 \lgs (typically
10--20\%) is not as significant as moving from 4 to 6 (a 30--70\% gain),
suggesting that using 6 \lgs would present a good trade-off between
cost and performance.

\begin{figure*}
(a)\includegraphics[width=0.45\linewidth]{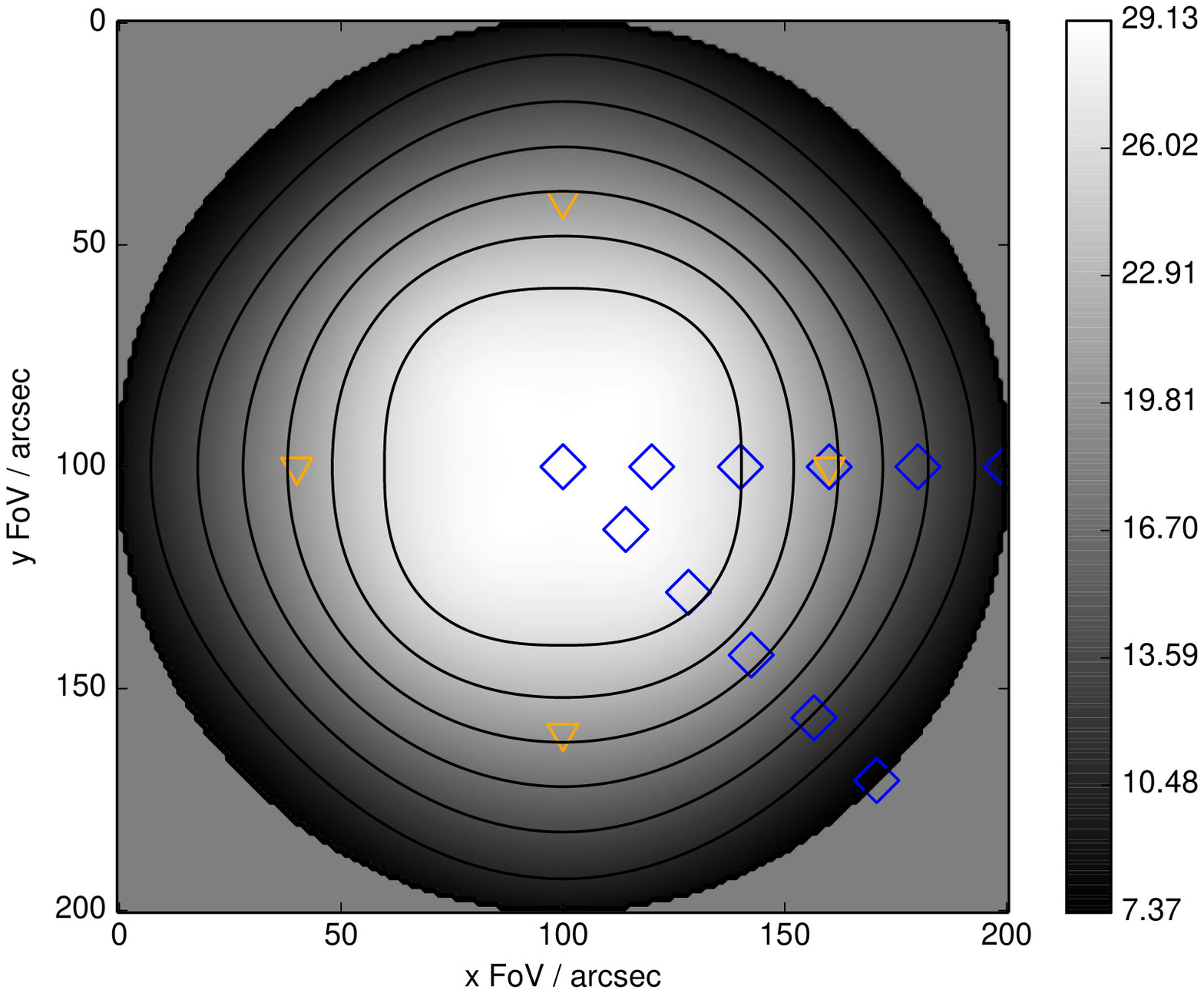}
(b)\includegraphics[width=0.45\linewidth]{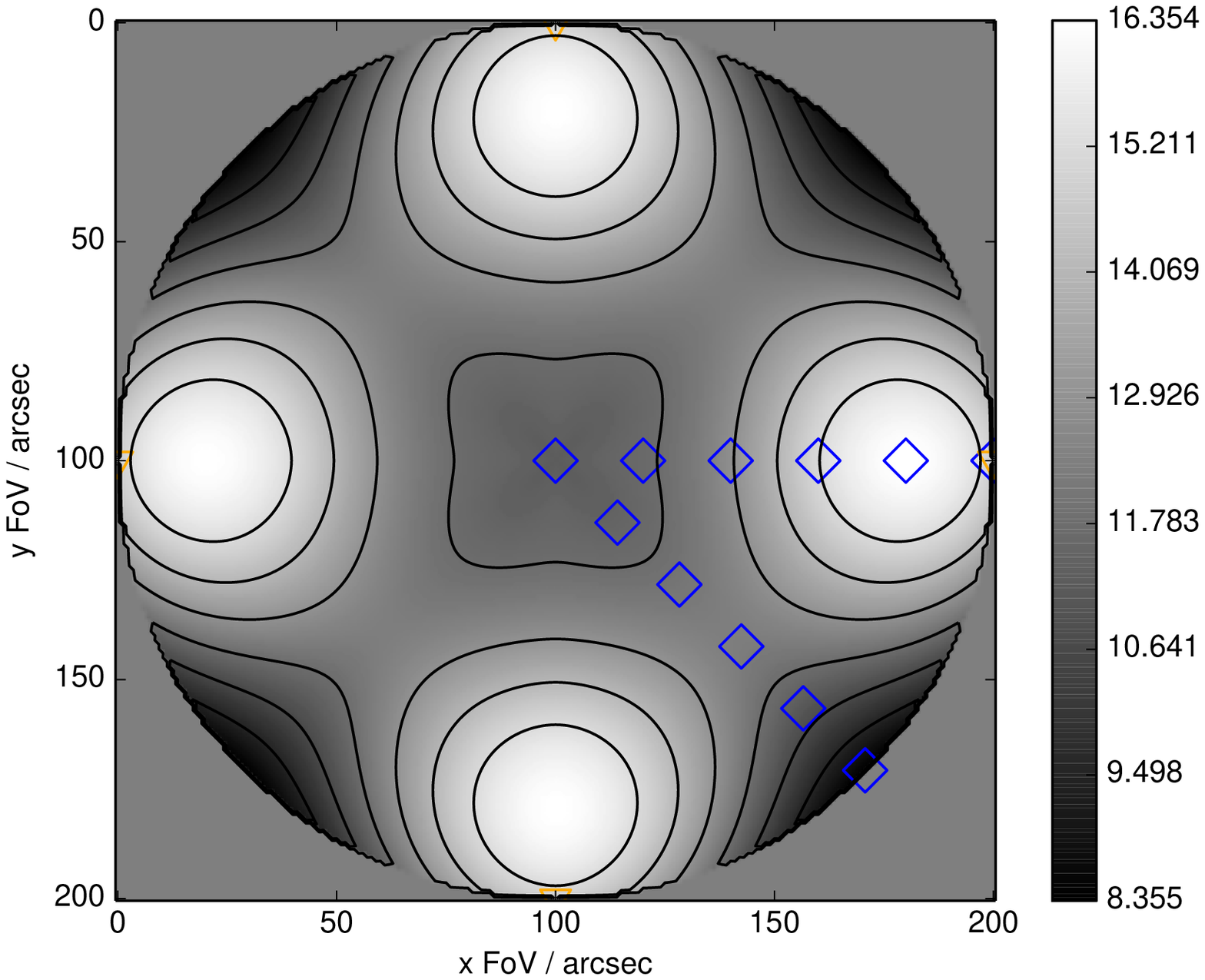}\\
(c)\includegraphics[width=0.45\linewidth]{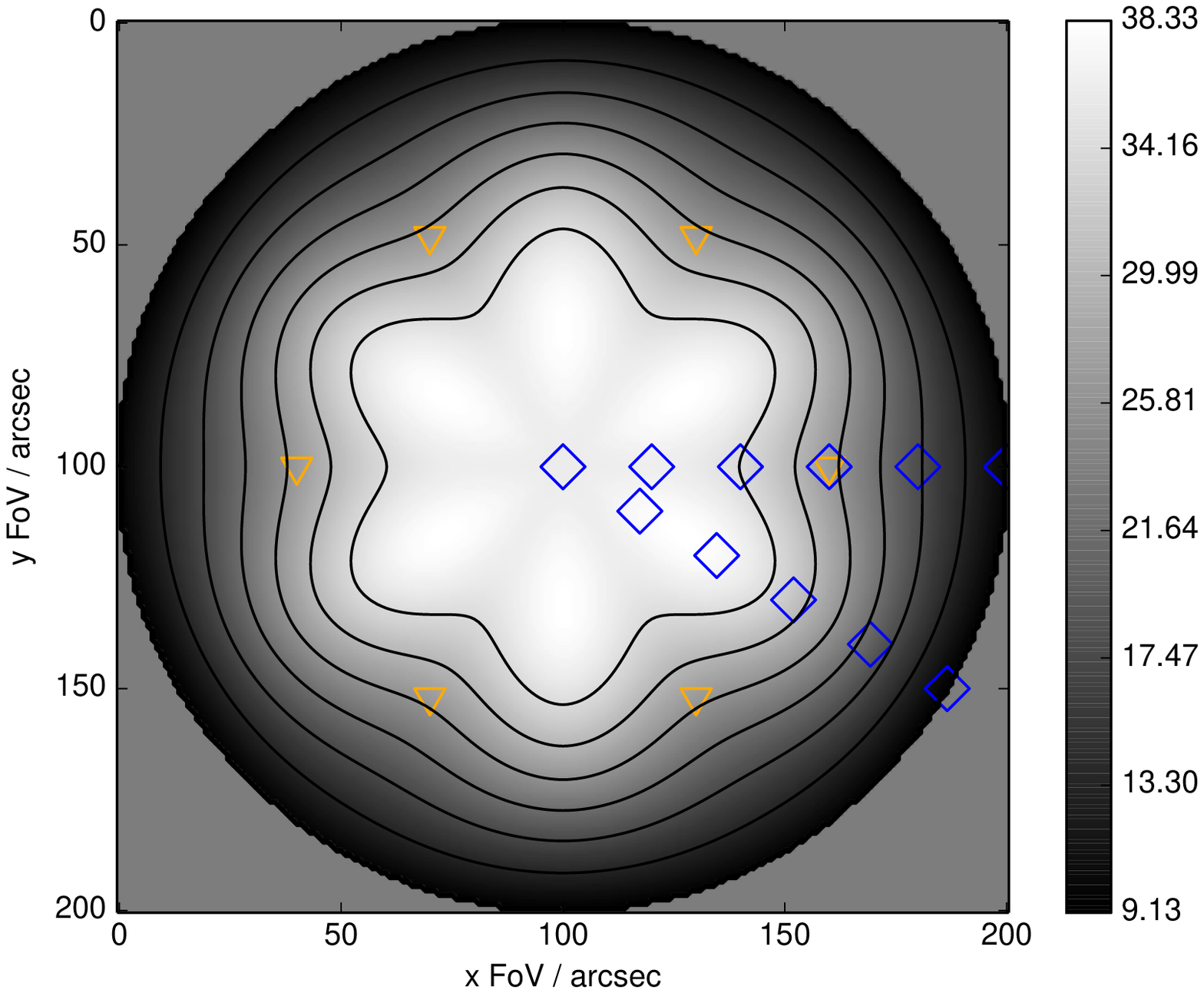}
(d)\includegraphics[width=0.45\linewidth]{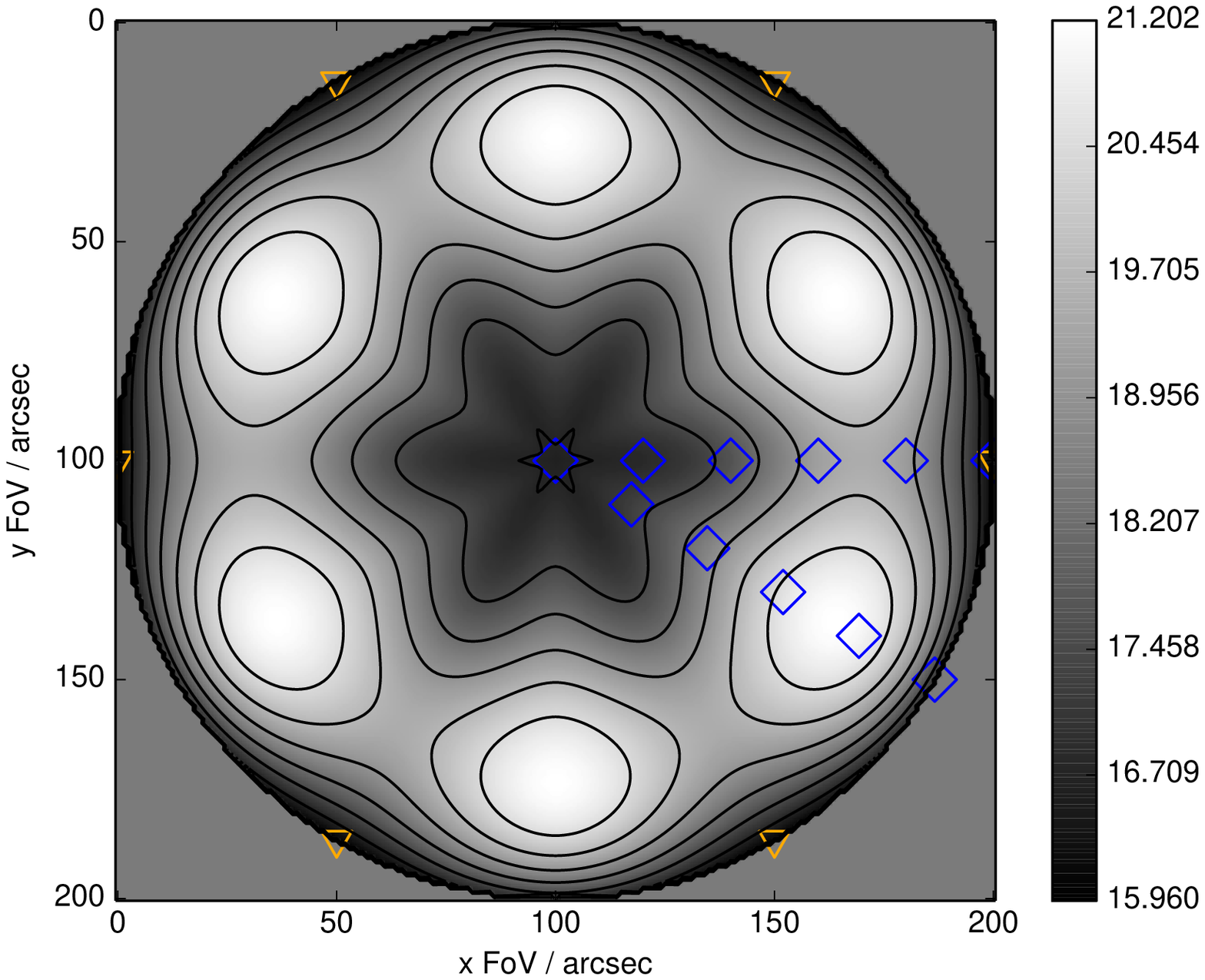}\\
(e)\includegraphics[width=0.45\linewidth]{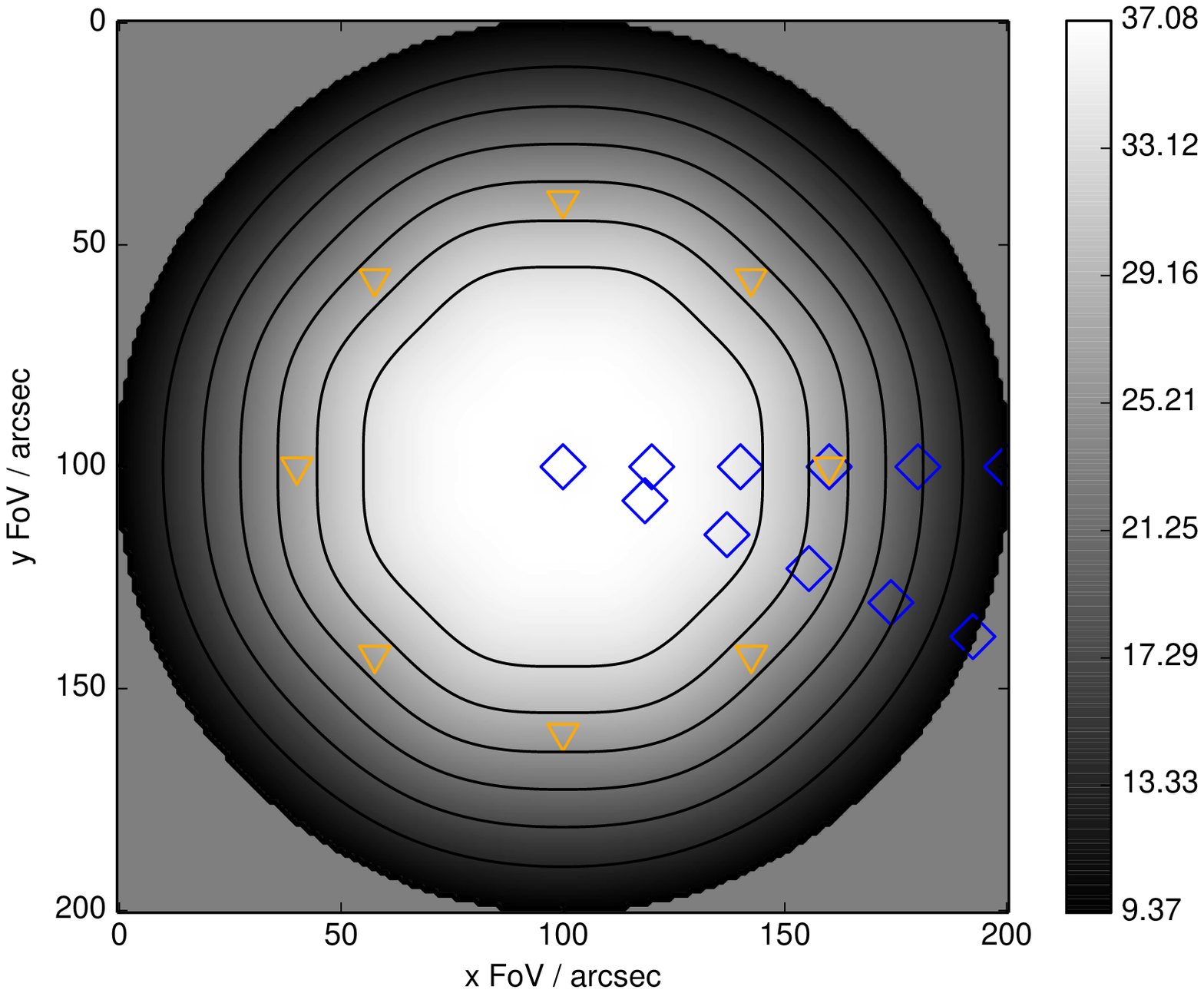}
(f)\includegraphics[width=0.45\linewidth]{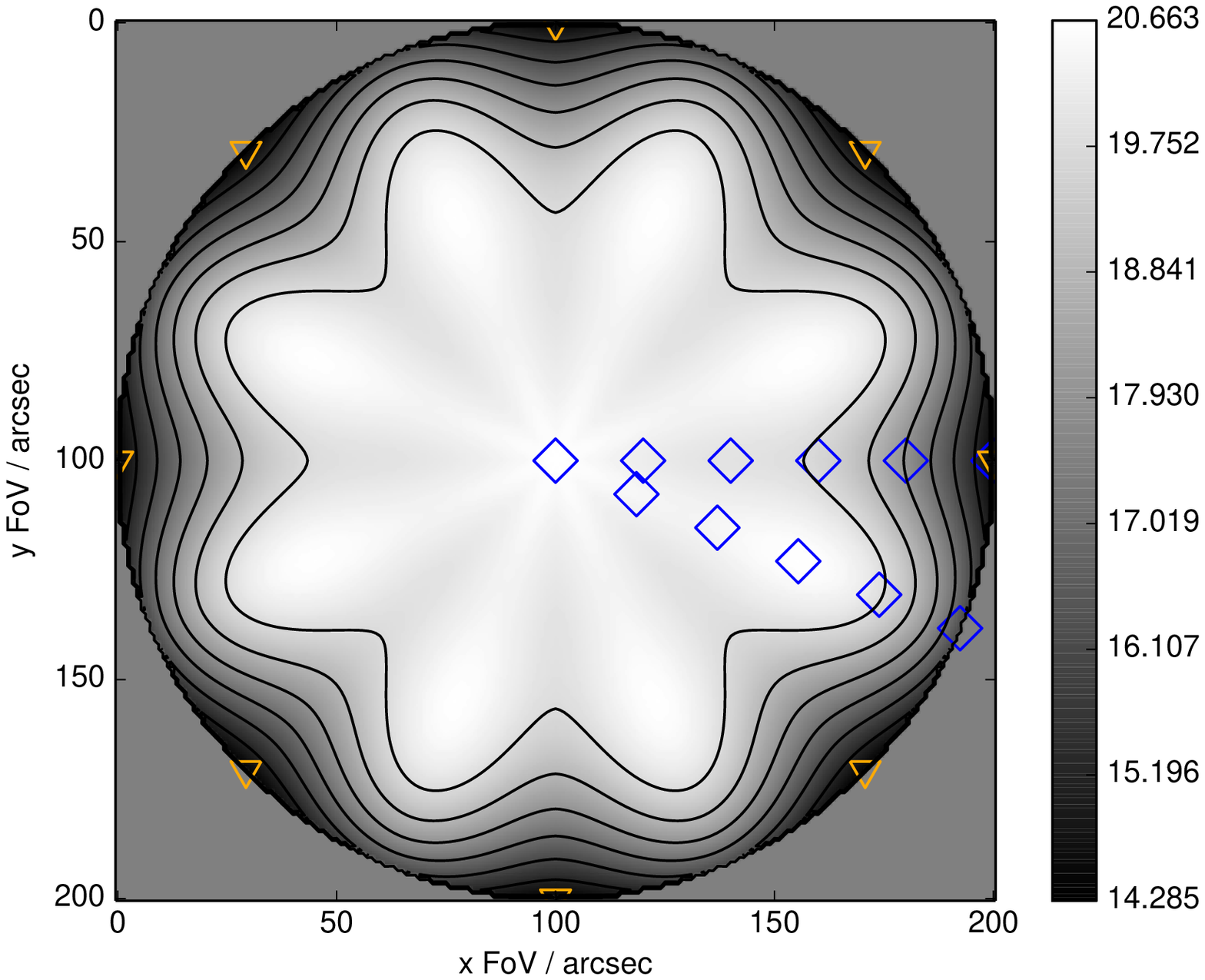}
\caption{A figure showing MCAO Strehl ratio across a 3.33 arcminute
  field of view, for: (top row) 4 LGS, (middle row) 6 LGS, (bottom
  row) 8 LGS, and for (left column) LGS on 2~arcminute diameter ring,
  (right column) LGS on 3.33~arcminute diameter ring.  The LGS
  positions are shown by orange triangles, and the science PSF
  sampling locations by blue diamonds.}
\label{fig:nlgsfield}
\end{figure*}

Field uniformity is not greatly affected by the number of \lgs, as
shown in Fig.~\ref{fig:nlgsfield}: the predicted \ao performance
remains reasonably constant across the central 2 arcminutes,
independent of the number of \lgs used (though with a uniformly lower
performance when fewer \lgs are used).  For comparison, we note that
we obtain a H-band Strehl ratio of about 50\% when modelling a \scao system
under identical conditions, with $74\times74$ sub-apertures, high
light level assumed, and an integrator control law, in
agreement with other studies \citep{2011aoel.confP..23C}.

\subsection{Dependence on DM conjugation}
\begin{figure}
\includegraphics[width=\linewidth]{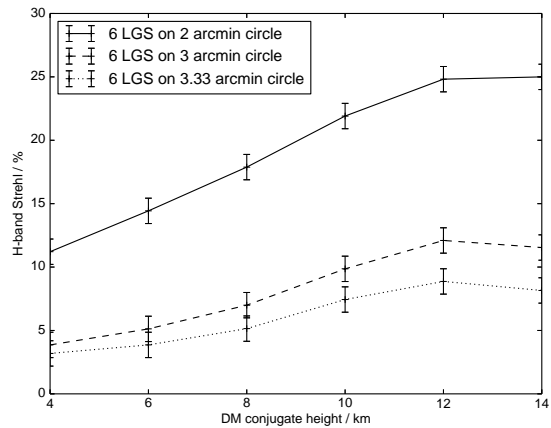}
\caption{A figure showing on-axis Strehl ratio as a function of upper DM
  conjugate height for a 2 DM MCAO system, with 6 LGSs.}
\label{fig:dmheight}
\end{figure}

\begin{figure}
\includegraphics[width=\linewidth]{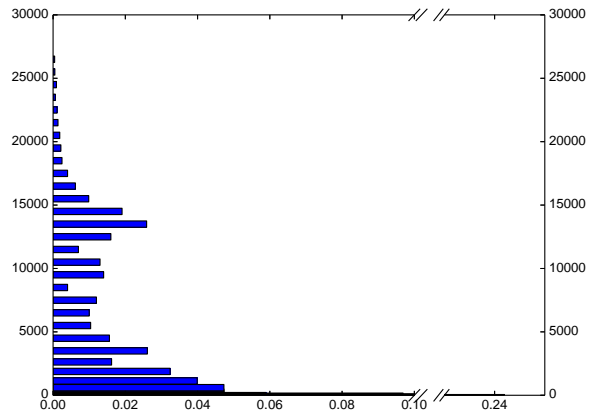}
\caption{The $C_n^2$ profile used in these simulations.}
\label{fig:cn2}
\end{figure}

Fig.~\ref{fig:dmheight} shows \mcao performance as a function of \dm
conjugation height, in the case of a 2 \dm system with 6 \lgss, with
the lower \dm conjugated to ground level.  For comparison, the
performance with 3 \dms conjugated at 0~km, 4~km and 12.7~km can be
taken from Fig.~\ref{fig:nlgs}.  We can see that best performance (for the
particular $C_n^2$ profile used, shown in Fig.~\ref{fig:cn2}) is
obtained with the upper \dm conjugated at 12~km, and that performance
with 3 \dms is significantly better than with 2.  When compared with
the $C_n^2$ profile (Fig.~\ref{fig:dmheight}), it is evident that
it is important to place the \dms according to where significant
turbulence strength lies.


\subsection{Dependence on DM actuator pitch}
\begin{figure}
\includegraphics[width=\linewidth]{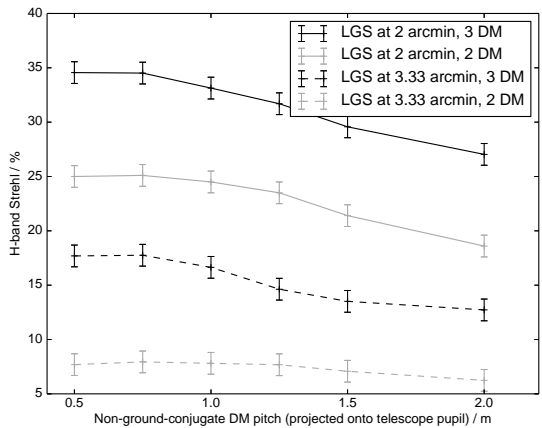}
\caption{A figure showing on-axis Strehl ratio as a function of above-ground
  layer DM actuator pitch, for 2 and 3 DMs, and for LGS asterism
  diameters of 2 and 3.33 arcminutes, as given in the legend.}
\label{fig:dmpitch}
\end{figure}

Fig.~\ref{fig:dmpitch} shows \mcao performance as a function of
above-ground \dm actuator pitch, for both the 2 \dm case (with the
upper \dm conjugated at 12~km), and the 3 \dm case.  The pitch of the
ground layer \dm is maintained at 52~cm.  It is evident that using a
1~m actuator pitch for above-ground layer \dms will lead to a small
performance degradation, compared to a smaller \dm pitch.  Using a
larger \dm actuator pitch results in predicted \ao performance falling
quickly.  A 0.75~m pitch delivers almost identical performance as a
0.5~m pitch.  Therefore, when performing a cost benefit analysis, it
would seem that the reduction in performance when using a 1~m actuator
pitch is acceptable given the cost reduction, but that further
increases in pitch would yield significant performance reduction.

\subsection{Conjugation height of ground layer DM}
\begin{figure}
\includegraphics[width=\linewidth]{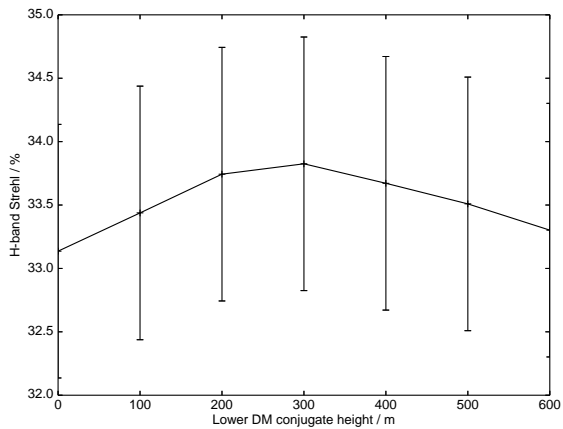}
\caption{A figure showing on-axis AO performance as a function of lowest DM
  conjugation height, for a 6~LGS (2~arcminute diameter spacing), 3~DM MCAO system.}
\label{fig:groundheight}
\end{figure}

The adaptive M4 mirror for the \eelt is not optically conjugated at
ground layer, but rather, a few hundred metres above ground.
Fig.~\ref{fig:groundheight} shows on-axis \ao system performance as a
function of ground-layer \dm conjugate height, and it is evident that
although there may be some variation in performance, this is small
over the range of likely conjugate heights, and so can be ignored.  We
do not take into account the differential conjugate height across the
\dm that results from the \eelt design (i.e.\ since the \dm is tilted,
one side has a lower optical conjugate than the other).  Instead, we
cover the range of conjugate heights.

\subsection{Performance variation with LGS pixel scale}
\begin{figure*}
(a)\includegraphics[width=0.45\linewidth]{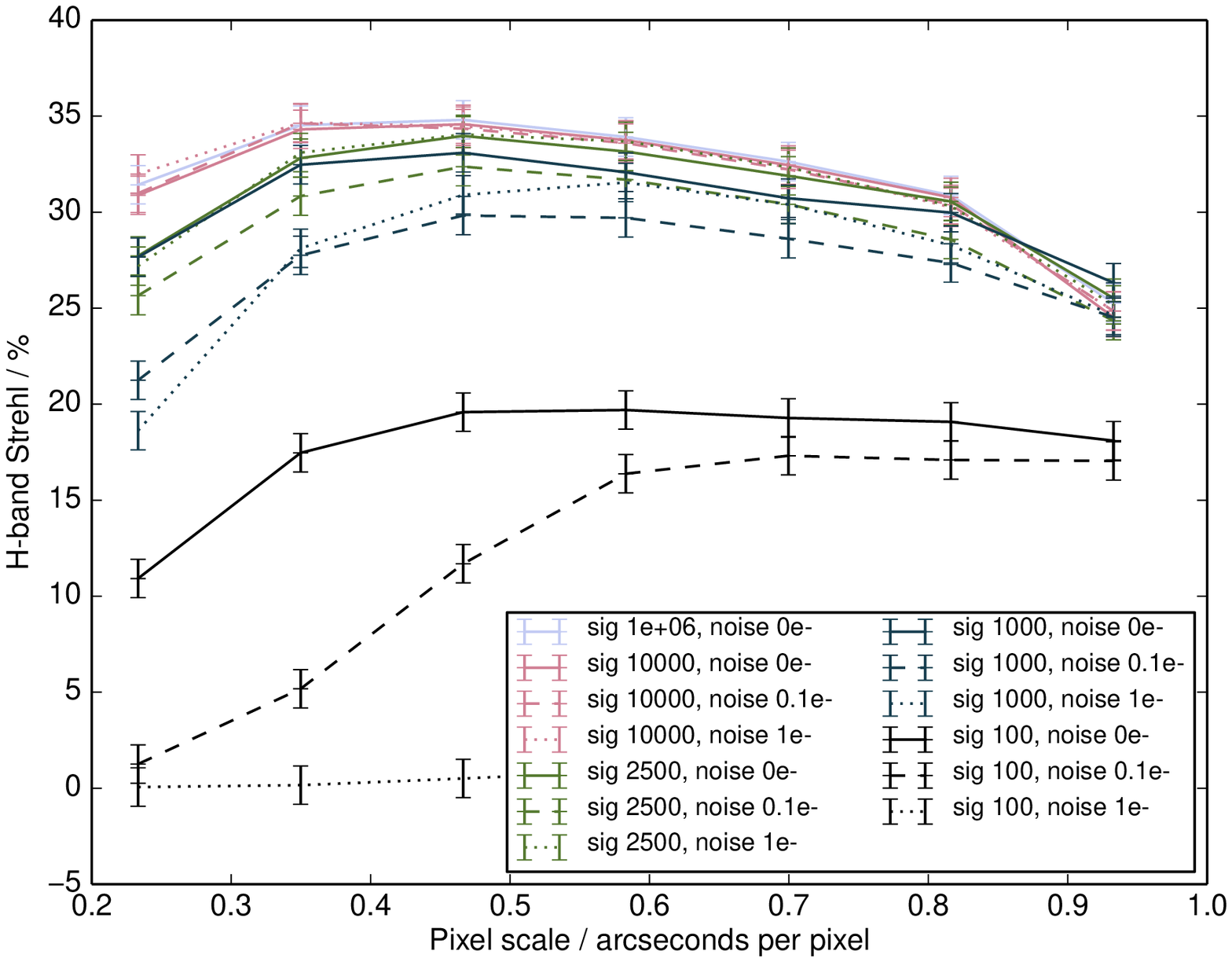}
(b)\includegraphics[width=0.45\linewidth]{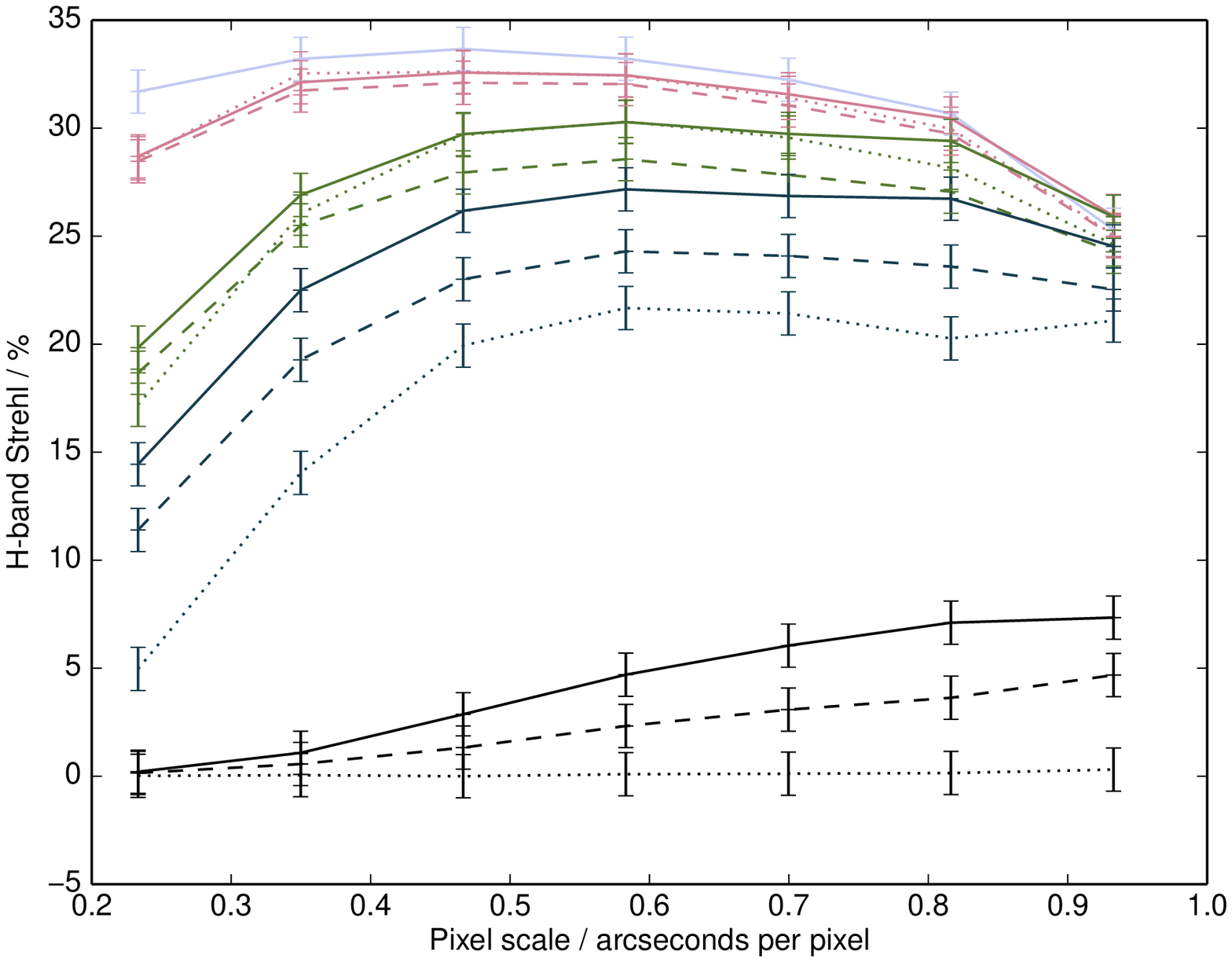}
(c)\includegraphics[width=0.45\linewidth]{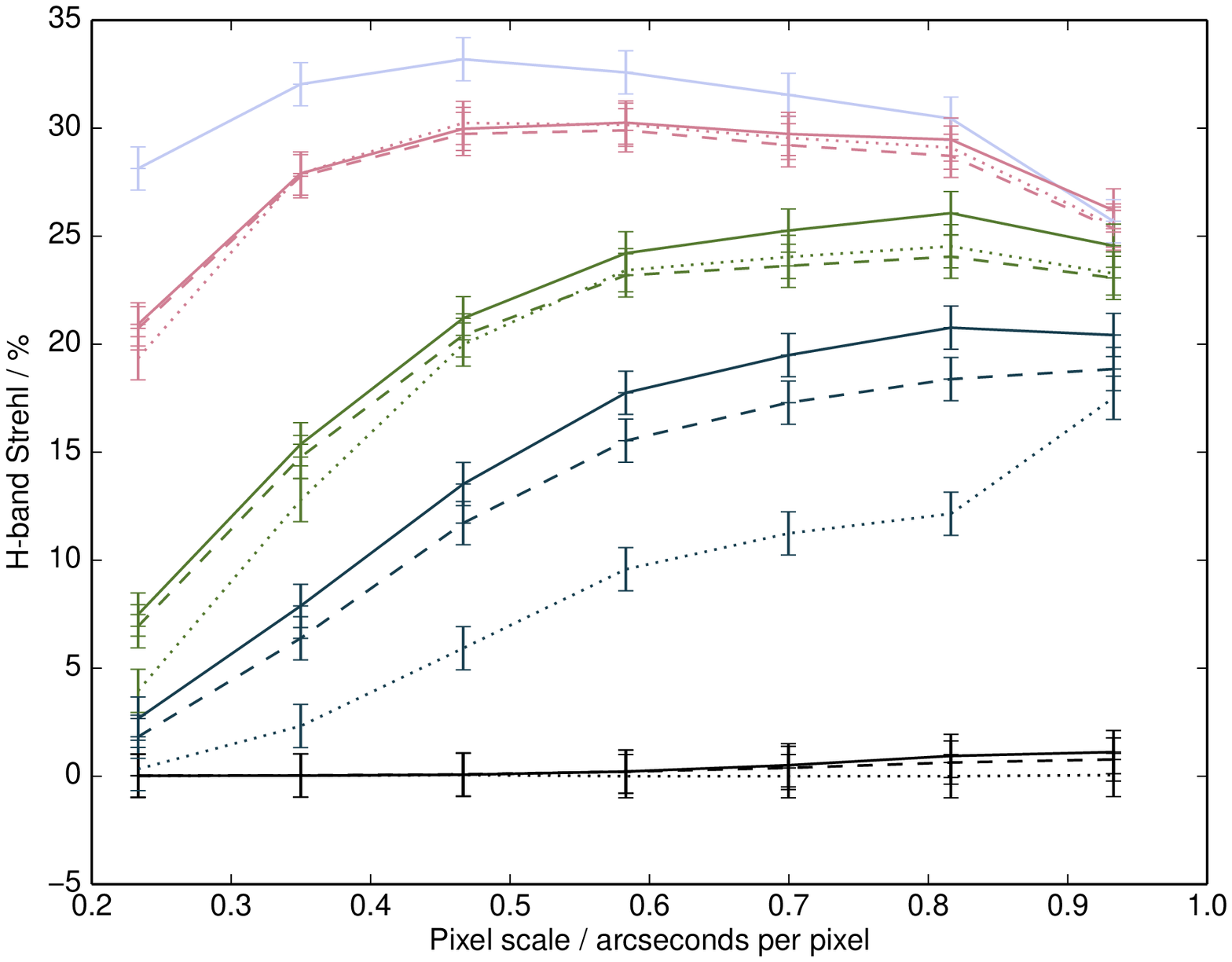}
(d)\includegraphics[width=0.45\linewidth]{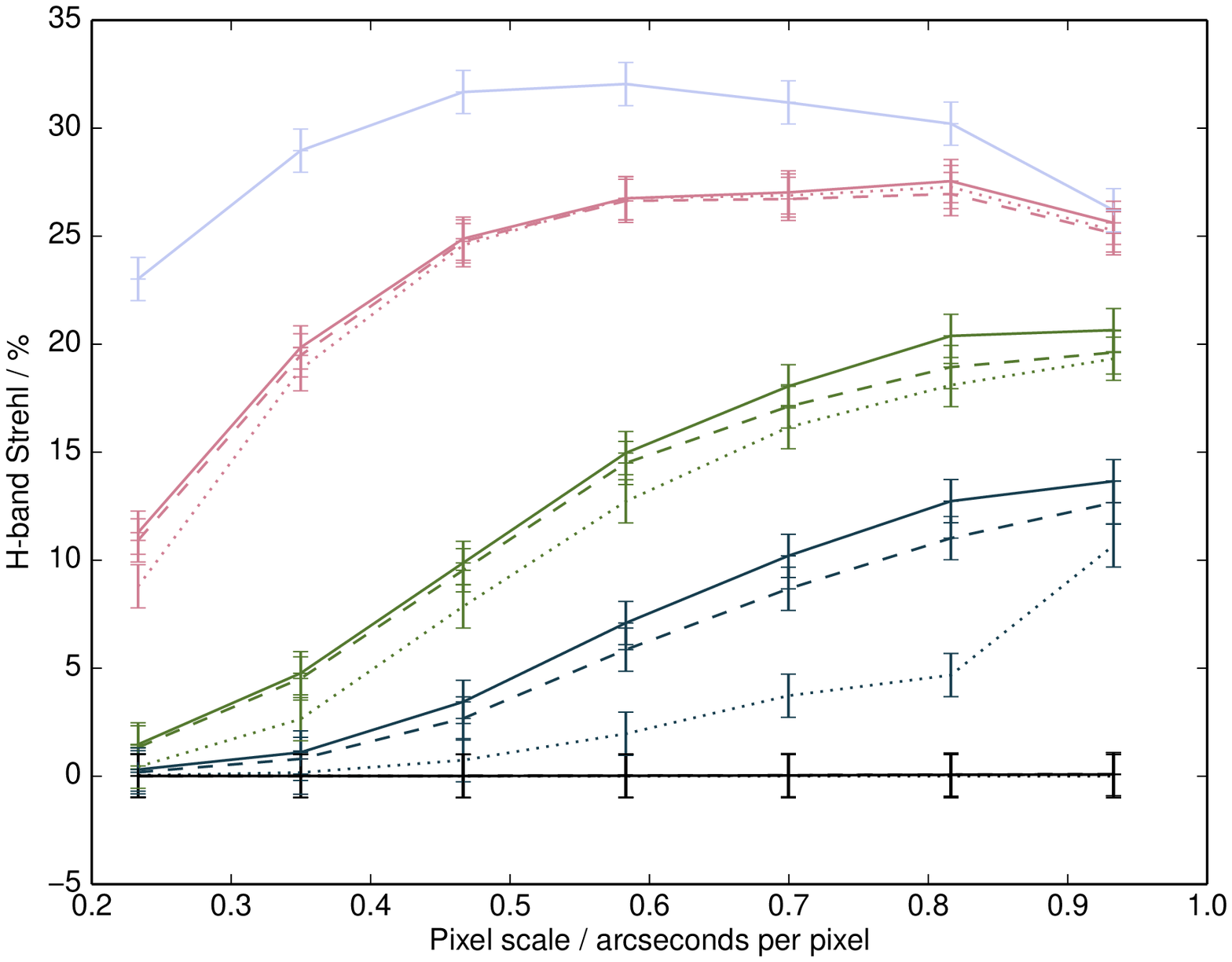}
\caption{A figure showing on-axis AO performance (H-band Strehl) as a
  function of wavefront sensor pixel scale, for (a) a 5~km sodium
  layer FWHM, (b) a 10~km sodium layer FWHM, (c) a 15~km sodium layer
  FWHM and (d) a 20~km sodium layer FWHM.  Different signal levels and
  readout noise levels are shown, as given by the legend in (a), using
  $16\times16$ pixel sub-apertures.  In summary, from dark to light
  represents increasing photon flux (sig, in photons per sub-aperture
  per frame), solid lines have no readout noise, dashed lines have
  0.1e- noise, and dotted lines have 1e- readout noise.}
\label{fig:sodiumdepth}
\end{figure*}

The default LGS pixel scale used throughout these simulations equates
to 0.23~arcsec/pixel, i.e.\ a \wfs field of view of 3.73~arcsec.  This
relatively small field of view is used to reduce computational
requirements.  As a result, our default sodium layer depth is also
relatively narrow, at 5~km \fwhm to avoid significant spot truncation.
Therefore, we investigate \ao performance as a function of both pixel
scale, and sodium layer depth, as shown in Fig.~\ref{fig:sodiumdepth}.
Since \ao performance is highly dependent on sub-aperture noise, which
is effected by signal level, detector characteristics, pixel scale and
sodium layer depth, we also investigate different noise levels here.
It can be seen that in all cases, there is an optimum pixel scale for
a given sodium \fwhm, signal level and readout noise level.  We
investigate detected photon fluxes of 100, 1000, 2500, 10000 and
10$^6$ photons per sub-aperture per frame, and consider readout noise
levels of 0, 0.1 and 1 photo-electrons rms (respresenting noiseless,
\emccd and \scmos technologies).  We note that a likely signal level
is between 1000--10000 photons per sub-aperture per frame, once
throughput losses have been taken into account.  

It can be seen that for the lowest signal levels with highest readout
noise and largest \lgs spots (as seen on the detector, i.e.\ large
sodium layer depth and small pixel scale), that \ao correction is very
poor, or fails.  In these cases, it would probably be possible to fine
tune the wavefront reconstruction algorithms, and use an optimal
sub-aperture processing algorithm to improve performance.  However, we
do not consider this here, as such optimisation is beyond the scope of
this paper.

At the likely signal levels of between 1000--10000 photons per
sub-aperture per frame, and a realistic sodium layer depth \fwhm of
10~km, these results suggest that a pixel scale of 0.6--0.7~arcseconds per
pixel is reasonable, being robust to changes in the sodium depth
(i.e.\ if the sodium layer depth changes, performance won't be
significantly affected).  If the sodium layer has a greater extent,
then a slightly larger pixel scale (say 0.8~arcseconds per pixel) may
be favourable for low signal-to-noise cases.

High signal-to-noise cases (high light level, low readout noise) are
seen to favour smaller pixel scales (around 0.4~arcseconds per pixel),
due to the increased \wfs sensitivity to spot motion (detectable phase
gradient resolution).  It can also be seen that at these light levels,
given the pixel scale is large enough, increasing the sodium layer
depth does not significantly impact \ao performance, i.e.\ the
performance curve has a broad peak.  However, at very low
pixel scales, significant truncation of \lgs spot images occurs,
resulting in reduced performance.  With large sodium layer depths, a
larger field of view can lead to increased performance, due to reduced
spot truncation, and hence higher sensitivity.  

We note for the default case (5~km depth, high light level, no noise),
that the variation in performance with pixel scale is small (about
10\% in Strehl).  Therefore, the results presented in the rest of this
paper using the default parameters are unlikely to be much different
when larger pixel scales are used.

\begin{figure}
\includegraphics[width=\linewidth]{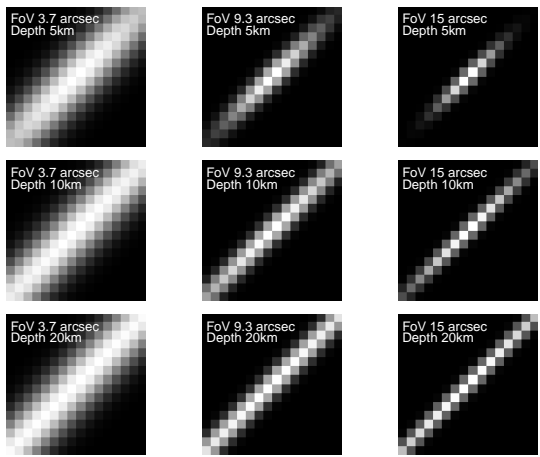}
\caption{A figure showing a single sub-aperture as simulated here,
  40~m from the laser launch location, for different sodium layer
  depths and pixel scales as given in the figure.}
\label{fig:elongation}
\end{figure}

Fig.~\ref{fig:elongation} shows the degree of \lgs elongation and
truncation for different pixel scales and sodium layer depths for
sub-apertures far from the laser launch locations (40~m away).  When
computing wavefront slopes for these sub-apertures, a conventional
centre of gravity algorithm has been used here, and we do not
explicitly take into account different slope noise characteristics
parallel and perpendicular to the elongation direction, nor do we
explicitly deal with bias introduced due to spot truncation, i.e.\ our
wavefront reconstruction is slightly pessimistic, and could be
improved upon in a separate study.

The \lgs spots are truncated at the edge of the sub-apertures, and we
assume that a field-stop is in place to prevent leakage between
sub-apertures.  

\subsection{LGS sub-aperture size}
\begin{figure}
\includegraphics[width=\linewidth]{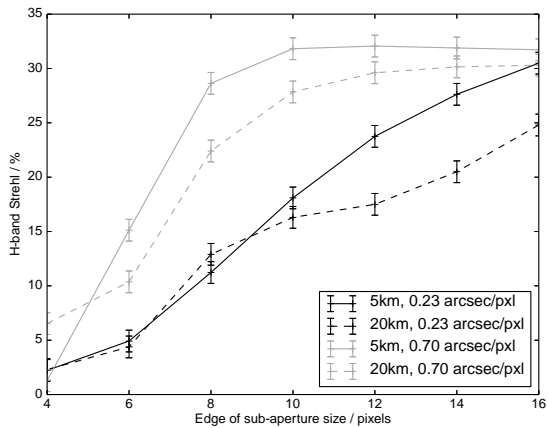}
\caption{A figure showing on-axis AO performance as a function of sub-aperture
  size, for different pixel scales and sodium layer depths (as given
  in the legend).}
\label{fig:subapsize}
\end{figure}

Larger sub-apertures require detectors with more pixels, increased
\rtcs computation power, more expensive detectors, and generally
result in more costly \ao systems.  We therefore explore \ao
performance as a function of sub-aperture size (in terms of pixel
count), in Fig.~\ref{fig:subapsize}.  It is evident here that larger
sub-apertures are favourable, primarily to avoid spot truncation.
With increased pixel scale (compressing the \lgs \psf into fewer
pixels), the number of pixels required (sub-aperture pixel size) can
be reduced with little performance loss.  However, with an increased
sodium layer depth (more elongated spots), resulting in increased spot
truncation for smaller sub-apertures, the pixel count cannot be
reduced without a more significant effect on performance, and larger
pixel count sub-apertures are favoured.

There is evidently a trade-off to be made, between sub-aperture size
and system cost.  We suggest that a minimum of $10\times10$ pixels per
sub-aperture would be appropriate for a pixel scale of 0.7~arcseconds
per pixel, with a total field of view of 7~arcseconds or more, though
this is highly dependent on actual sodium layer profile.  We note that
this is a minimum requirement, and that greater performance would be
achieved using a larger number of pixels particularly when the sodium
layer depth is more extensive, provided that readout noise does not
dominate.

\subsection{Operation with NGS}
\begin{figure}
\includegraphics[width=\linewidth]{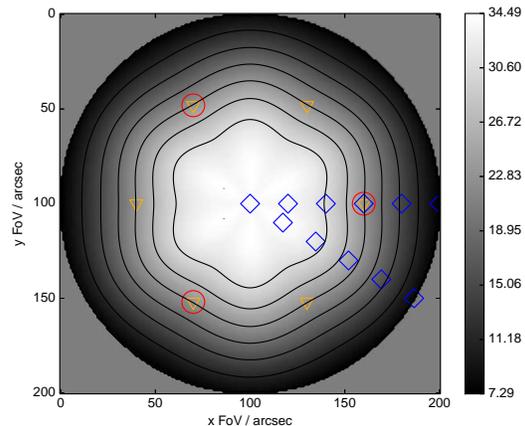}
\caption{A figure showing Strehl across a 3.33 arcminute field of view
  for tip-tilt correction performed using 3 NGS, and higher order
  correction using 6 LGS.  This can be compared directly with
  Fig.~\ref{fig:nlgsfield}(c) which shows LGS correction only (i.e.\ tip-tilt
  correction is performed using the LGS).  The LGS positions are shown
by orange triangles, the NGS positions by red circles, and the science PSF sampling
positions by blue diamonds.}
\label{fig:ngstt}
\end{figure}

Fig.~\ref{fig:ngstt} shows predicted \ao performance as a function of
distance from the on-axis direction, for a system using 3 \ngs for
tip-tilt correction, and 6 \lgs for higher order correction.  This can
be compared directly with Fig.~\ref{fig:nlgsfield}(c), which uses only
\lgs (that are used unphysically for tip-tilt correction).
Performance is very similar in both cases (though not identical),
which confirms that the simplification made when using only \lgss is
able to provide a reasonably reliable performance estimate.  As stated
previously, we note that actual performance will depend somewhat on
\ngs asterism shape and availability of suitably bright targets,
though this is beyond the scope of the study presented here.  In the
case presented here, the 3 \ngss were in an asterism equally placed
around a 2~arcminute diameter circle, as were the 6 \lgs.

\subsection{Comparisons with other simulation results}
A direct comparison with other previous simulation results is
difficult, due to differences in atmospheric models, science
wavelength, numbers of sub-apertures, telescope size and other
modelling uncertainties and differences.  However, verification of
performance trends is possible.  We find that our estimated
performance for basic models is broadly similar to other \elt
Monte-Carlo models
\citep{miskaltao,2011aoel.confE..63T,2010aoel.confE2013F}, and
slightly pessimistic when compared with analytical models of \ao
performance \citep{2008JOSAA..26..219N}, as expected.  The
consideration of \dm conjugation is largely independent of telescope
diameter, and our results are similar to those given by
\citet{2003A&A...404.1165F,2000SPIE.4007.1032F}.  Similarly, a study
of \lgs pixel count was carried out by \citet{2011aoel.confE..67V}.  A
combined study of pixel scale, sodium layer depth and \lgs signal
level is new here.

\section{Conclusions}
We have performed \ao performance modelling for a \mcao system on the
\eelt, using a Monte-Carlo, end-to-end \ao simulation code, and have
looked at number of \lgss, \dm configuration, \lgs pixel scale, and
sodium layer depth.  We find that using 6 \lgs seems to be a good
compromise between performance and cost.  The use of 3 \dms, rather
than 2 provides a significant performance advantage, though it is
possible to reduce the actuator pitch of these \dms to below that of
the \wfss, without significant performance loss, hence reducing system
cost.  We find that the ideal pixel scale and \wfs field of view
depends on sodium layer profile, and suggest that a field of view
should be chosen that is sufficient to encompass all likely sodium
layer profile depths.  A pixel scale of at least
0.7~arcseconds per pixel is necessary, and at least $10\times10$
pixels per sub-aperture should be used, for the simplified Gaussian
profiles used here.  We note that this is likely to lead to spot
truncation, and there is a trade-off between truncation and
sensitivity. We also find that, as expected, larger sub-apertures (in
terms of pixel count) offer better performance as this reduces
clipping of the elongated \lgs spots.

\section*{Acknowledgements}
This work is funded by the UK Science and Technology Facilities
Council, grant ST/K003569/1, and a consolidated grant ST/L00075X/1.
Helpful comments from Tim Morris are acknowledged.
\bibliographystyle{mn2e}

\bibliography{mybib}

\begin{thebibliography}{}

\bibitem[\protect\citeauthoryear{{Arcidiacono}, {Schreiber}, {Bregoli},
  {Diolaiti}, {Foppiani}, {Cosentino}, {Lombini}, {Butler} \&
  {Ciliegi}}{{Arcidiacono} et~al.}{2014}]{2014SPIE.9148E..6FA}
{Arcidiacono} C.,  {Schreiber} L.,  {Bregoli} G.,  {Diolaiti} E.,  {Foppiani}
  I.,  {Cosentino} G.,  {Lombini} M.,  {Butler} R.~C.,    {Ciliegi} P.,  2014,
  in Society of Photo-Optical Instrumentation Engineers (SPIE) Conference
  Series Vol.~9148 of Society of Photo-Optical Instrumentation Engineers (SPIE)
  Conference Series, {End to end numerical simulations of the MAORY
  multiconjugate adaptive optics system}.
p.~6

\bibitem[\protect\citeauthoryear{{Babcock}}{{Babcock}}{1953}]{adaptiveoptics}
{Babcock} H.~W.,  1953, \pasp, 65, 229

\bibitem[\protect\citeauthoryear{{Basden}}{{Basden}}{2015}]{basden19}
{Basden} A.,  2015, JATIS, 1(3), 039002

\bibitem[\protect\citeauthoryear{{Basden}}{{Basden}}{2014}]{basden15}
{Basden} A.~G.,  2014, \mnras, 440, 577

\bibitem[\protect\citeauthoryear{{Basden}, {Butterley}, {Myers} \&
  {Wilson}}{{Basden} et~al.}{2007}]{basden5}
{Basden} A.~G.,  {Butterley} T.,  {Myers} R.~M.,    {Wilson} R.~W.,  2007,
  Appl.\ Optics, 46, 1089

\bibitem[\protect\citeauthoryear{{Basden}, {Evans} \& {Morris}}{{Basden}
  et~al.}{2014}]{basden17}
{Basden} A.~G.,  {Evans} C.~J.,    {Morris} T.~J.,  2014, \mnras, 445, 4008

\bibitem[\protect\citeauthoryear{{Basden} \& {Myers}}{{Basden} \&
  {Myers}}{2012}]{basden11}
{Basden} A.~G.,  {Myers} R.~M.,  2012, \mnras, 424, 1483

\bibitem[\protect\citeauthoryear{{Bonaccini Calia}}{{Bonaccini
  Calia}}{2015}]{caliaPrivate}
{Bonaccini Calia} D., , 2015, private communication, presentation at EWASS 2015

\bibitem[\protect\citeauthoryear{{Cl{\'e}net}, {Bernardi}, {Chapron},
  {Gendron}, {Rousset}, {Hubert}, {Davies}, {Thiel} \& {Tromp}}{{Cl{\'e}net}
  et~al.}{2011}]{2011aoel.confP..23C}
{Cl{\'e}net} Y.,  {Bernardi} P.,  {Chapron} F.,  {Gendron} E.,  {Rousset} G.,
  {Hubert} Z.,  {Davies} R.,  {Thiel} M.,    {Tromp} N.,  2011, in Second
  International Conference on Adaptive Optics for Extremely Large Telescopes.
  Online at <A
  href=''http://ao4elt2.lesia.obspm.fr''>http://ao4elt2.lesia.obspm.fr</A>,
  id.P23 {The SCAO module of the E-ELT adaptive optics imaging camera MICADO}.
p.~23P

\bibitem[\protect\citeauthoryear{Ellerbroek, Gilles \& Vogel}{Ellerbroek
  et~al.}{2003}]{map}
Ellerbroek B.,  Gilles L.,    Vogel C.,  2003, Appl.\ Optics, 42, 4811

\bibitem[\protect\citeauthoryear{{Femen{\'{\i}}a} \&
  {Devaney}}{{Femen{\'{\i}}a} \& {Devaney}}{2003}]{2003A&A...404.1165F}
{Femen{\'{\i}}a} B.,  {Devaney} N.,  2003, \aap, 404, 1165

\bibitem[\protect\citeauthoryear{{Flicker}, {Rigaut} \& {Ellerbroek}}{{Flicker}
  et~al.}{2000}]{2000SPIE.4007.1032F}
{Flicker} R.,  {Rigaut} F.~J.,    {Ellerbroek} B.~L.,  2000, in {Wizinowich}
  P.~L.,  ed., Adaptive Optical Systems Technology Vol.~4007 of Society of
  Photo-Optical Instrumentation Engineers (SPIE) Conference Series, {Comparison
  of multiconjugate adaptive optics configurations and control algorithms for
  the Gemini South 8-m telescope}.
pp 1032--1043

\bibitem[\protect\citeauthoryear{{Foppiani}, {Diolaiti}, {Lombini},
  {Baruffolo}, {Biliotti}, {Bregoli}, {Cosentino}, {Delabre}, {Marchetti},
  {Schreiber}, {Conan}, {D'Odorico} \& {Hubin}}{{Foppiani}
  et~al.}{2010}]{2010aoel.confE2013F}
{Foppiani} I.,  {Diolaiti} E.,  {Lombini} M.,  {Baruffolo} A.,  {Biliotti} V.,
  {Bregoli} G.,  {Cosentino} G.,  {Delabre} B.,  {Marchetti} E.,  {Schreiber}
  L.,  {Conan} J.-M.,  {D'Odorico} S.,    {Hubin} N.,  2010, in Adaptative
  Optics for Extremely Large Telescopes {MCAO for the E-ELT: preliminary design
  overview of the MAORY module}.
p.~2013

\bibitem[\protect\citeauthoryear{{Foy} \& {Labeyrie}}{{Foy} \&
  {Labeyrie}}{1985}]{laserguidestar}
{Foy} R.,  {Labeyrie} A.,  1985, \aap, 152, L29

\bibitem[\protect\citeauthoryear{Johns}{Johns}{2008}]{gmt}
Johns M.,  2008, in Extremely Large Telescopes: Which Wavelengths? Retirement
  Symposium for Arne Ardeberg Vol.~6986, The giant magellan telescope (gmt).
pp 698603--698603--12

\bibitem[\protect\citeauthoryear{Le~Louarn, Clare, Béchet \& Tallon}{Le~Louarn
  et~al.}{2012}]{miskaltao}
Le~Louarn M.,  Clare R.,  Béchet C.,    Tallon M.,  2012 Vol.~8447,
  Simulations of adaptive optics systems for the e-elt.
pp 84475D--84475D--7

\bibitem[\protect\citeauthoryear{{Neichel}, {Fusco} \& {Conan}}{{Neichel}
  et~al.}{2008}]{2008JOSAA..26..219N}
{Neichel} B.,  {Fusco} T.,    {Conan} J.-M.,  2008, Journal of the Optical
  Society of America A, 26, 219

\bibitem[\protect\citeauthoryear{{Nelson} \& {Sanders}}{{Nelson} \&
  {Sanders}}{2008}]{tmt}
{Nelson} J.,  {Sanders} G.~H.,  2008, in Society of Photo-Optical
  Instrumentation Engineers (SPIE) Conference Series Vol.~7012 of Society of
  Photo-Optical Instrumentation Engineers (SPIE) Conference Series, {The status
  of the Thirty Meter Telescope project}.
pp 70121A--70121A--18

\bibitem[\protect\citeauthoryear{{Otarola}, {Neichel}, {Wang}, {Boyer},
  {Ellerbroek} \& {Rigaut}}{{Otarola} et~al.}{2013}]{2013aoel.confE..58O}
{Otarola} A.,  {Neichel} B.,  {Wang} L.,  {Boyer} C.,  {Ellerbroek} B.,
  {Rigaut} F.,  2013, in {Esposito} S.,  {Fini} L.,  eds, Proceedings of the
  Third AO4ELT Conference {Analysis of fratricide effect observed with GeMS and
  its relevance for large aperture astronomical telescopes}.
p.~58

\bibitem[\protect\citeauthoryear{{Rigaut}, {Neichel}, {Boccas}, {d'Orgeville},
  {Arriagada}, {Fesquet}, {Diggs}, {Marchant}, {Gausach}, {Rambold}, {Luhrs},
  {Walker}, {Carrasco-Damele}, {Edwards}, {Pessev} \& {Galvez}}{{Rigaut}
  et~al.}{2012}]{2012SPIE.8447E..0IRshort}
{Rigaut} F.,  {Neichel} B.,  {Boccas} M.,  {d'Orgeville} C.,  {Arriagada} G.,
  {Fesquet} V.,  {Diggs} S.~J.,  {Marchant} C.,  {Gausach} G.,  {Rambold}
  W.~N.,  {Luhrs} J.,  {Walker} S.,  {Carrasco-Damele} E.~R.,  {Edwards} M.~L.,
   {Pessev} P.,    {Galvez} R.~L.,  2012, in Society of Photo-Optical
  Instrumentation Engineers (SPIE) Conference Series Vol.~8447 of Society of
  Photo-Optical Instrumentation Engineers (SPIE) Conference Series, {GeMS:
  first on-sky results}

\bibitem[\protect\citeauthoryear{{Sarazin}, {Le Louarn}, {Ascenso}, {Lombardi}
  \& {Navarrete}}{{Sarazin} et~al.}{2013}]{35layer}
{Sarazin} M.,  {Le Louarn} M.,  {Ascenso} J.,  {Lombardi} G.,    {Navarrete}
  J.,  2013, in Adaptative Optics for Extremely Large Telescopes 3 {Defining
  reference turbulence profiles for E-ELT AO performance simulations}

\bibitem[\protect\citeauthoryear{{Spyromilio}, {Comer{\'o}n}, {D'Odorico},
  {Kissler-Patig} \& {Gilmozzi}}{{Spyromilio} et~al.}{2008}]{eelt}
{Spyromilio} J.,  {Comer{\'o}n} F.,  {D'Odorico} S.,  {Kissler-Patig} M.,
  {Gilmozzi} R.,  2008, The Messenger, 133, 2

\bibitem[\protect\citeauthoryear{{Tallon}, {B{\'e}chet}, {Tallon-Bosc}, {Le
  Louarn}, {Thi{\'e}baut}, {Clare} \& {Marchetti}}{{Tallon}
  et~al.}{2011}]{2011aoel.confE..63T}
{Tallon} M.,  {B{\'e}chet} C.,  {Tallon-Bosc} I.,  {Le Louarn} M.,
  {Thi{\'e}baut} {\'E}.,  {Clare} R.,    {Marchetti} E.,  2011, in Second
  International Conference on Adaptive Optics for Extremely Large Telescopes.
  Online at <A
  href=''http://ao4elt2.lesia.obspm.fr''>http://ao4elt2.lesia.obspm.fr</A>,
  id.63 {Performance of MCAO on the E-ELT using the Fractal Iterative Method
  for fast atmospheric tomography}.
p.~63

\bibitem[\protect\citeauthoryear{{van Dam}, {Conan}, {Bouchez} \&
  {Espeland}}{{van Dam} et~al.}{2011}]{2011aoel.confE..67V}
{van Dam} M.~A.,  {Conan} R.,  {Bouchez} A.~H.,    {Espeland} B.,  2011, in
  Second International Conference on Adaptive Optics for Extremely Large
  Telescopes. Online at <A
  href=''http://ao4elt2.lesia.obspm.fr''>http://ao4elt2.lesia.obspm.fr</A>,
  id.67 {Aberrations induced by side-projected laser guide stars in laser
  tomography adaptive optics systems}.
p.~67

\end{thebibliography}
\bsp

\end{document}